\documentclass[doublecol]{epl2}
\usepackage[latin9]{inputenc}
\usepackage[active]{srcltx}
\usepackage{amssymb}
\usepackage{graphicx}
\usepackage{amsmath}
\usepackage{textcomp}
\usepackage{url}

\makeatletter
 
\usepackage{amsmath}

\title{Dynamic Clustering in Suspension of Motile Bacteria}
\shorttitle{Dynamic Clustering in Suspension of Motile Bacteria} 

\author{Xiao Chen\inst{1} \and Xiang Yang \inst{1} \and Mingcheng Yang \inst{2}  \and H. P. Zhang\inst{1,3}\thanks{E-mail:\email{hepeng\_zhang@sjtu.edu.cn}}}
\shortauthor{Xiao Chen\etal}

\institute{                    
  \inst{1} Department of Physics and astronomy and Institute of Natural Sciences, Shanghai Jiao Tong University, China \\ 
  \inst{2} Beijing National Laboratory for Condensed Matter Physics and Key Laboratory of Soft Matter Physics, Institute of Physics, Chinese Academy of Sciences, China \\
  \inst{3} Collaborative Innovation Center of Advanced Microstructures, Nanjing, China  
}


\abstract{Bacteria suspension exhibits a wide range of collective phenomena
arsing from interactions between individual cells. Here we show 
\textit{Serratia marcescens} cells near an air-liquid interface spontaneously
aggregate into dynamic clusters through surface-mediated hydrodynamic
interactions. These long-lived clusters translate randomly and rotate
in the counter-clockwise direction; they continuously evolve, merge
with others and split into smaller ones. Measurements indicate that long-ranged hydrodynamic interactions have strong influences on cluster properties. Bacterial clusters change material and fluid transport near the interface and hence may have environmental and biological consequences.}

\makeatother

\begin{document}
\maketitle

Active systems are composed of self-propelled particles that can produce
motion by taking in and dissipating energy\cite{Ramaswamy2010,vicsekreview,RevModPhysMarchetti,Elgeti2015}.
Examples exist at different length scales, from bacteria suspension
\cite{Wu2000,Dombrowski2004,Sokolov2007,Zhang2010a,Chen2012,Gachelin2013}
to flocks of birds\cite{Ballerini2008,Nagy2010,Cavagna2014}. Being
far from thermal equilibrium, active systems are not subject to thermodynamic
constraints, such as detailed balance or fluctuation-dissipation theorem
\cite{Tailleur2008,Cates2012,Cates2013}. This renders the physics
of active systems much richer than that of thermal systems. For example,
collective motion with extended spatio-temporal coherence has been
reported in many active systems \cite{Wu2000,Dombrowski2004,Sokolov2007,Narayan2007,Ballerini2008,Zhang2009,Zhang2010a,Nagy2010,Chen2012,Bricard2013,Cavagna2014}.
Such coherent motion can arise from local interactions that align
a particle's motion with its neighbors through biological coordination
\cite{Ballerini2008,Nagy2010} or physical interactions \cite{Narayan2007,Bricard2013}.

Active systems without alignment interactions also exhibit interesting
collective behavior. Theoretical models have shown that systems with
a density-dependent motility phase separate into dense dynamic clusters
and a dilute gas phase\cite{Tailleur2008,Cates2012,cates2015}. Numerical
simulations of repulsive self-propelled disks confirmed the theoretical
prediction of phase separation\cite{Fily2012,Redner2013}. Effects
of motility, attractive interaction, and hydrodynamic forces have
been extensively explored in simulations\cite{Mognetti,Zoettl2014,Furukawa2014,Matas-Navarro2014}.
On the experimental side, dynamic clusters have been observed in Janus
particles (platinum-coated \cite{Palacci2010} and Carbon-coated\cite{Buttinoni2013})
and colloidal particles with an embedded hematic cube \cite{Palacci2013}.
Schwarz-Linek \textit{et al} observed clusters of motile bacteria
when they added polymers to bacteria suspension to induce depletion
attraction between bacteria \cite{Schwarz-Linek2012}. In a very recent
paper \cite{Petroff}, Petroff \textit{et al} reported that \textit{Thiovulum
majus }bacteria form two-dimensional crystals near a liquid-solid
interface. Understanding the origins and properties of these dynamic
clusters may provide new insights into emergent behaviors of active
matters and open up possibilities to build novel materials \cite{Cates2012}.

In this letter, we report experimental results for a new type of bacterial
clusters formed near an air-liquid interface in a pure suspension
without depletant agents. Fluid dynamic calculation and flow visualization
are used to show surface-mediated hydrodynamic interactions can explain
the formation of these clusters. We further quantify the statistical
and dynamic properties of bacterial clusters and show long-ranged
hydrodynamic forces have important influences on cluster properties.
We conclude with discussions on related research and on possible technological
and environmental implications of our work.

\begin{figure}
\centering{}\includegraphics[width=7.5cm]{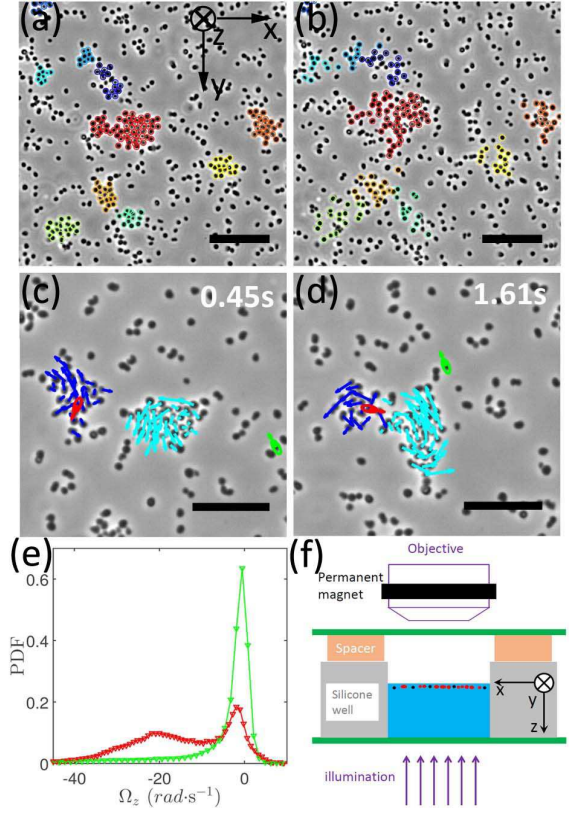}\protect\caption{(color online). (a) Dynamic clusters observed in a sample with bacterial
density $\phi=0.058\mbox{}\mu\mbox{m}^{-2}$. (b) Clusters dissolve
after flagellar motor is damaged by strong light. The false colors
in (a-b) show the time evolution of bacteria originally belonging
to different clusters; image in (b) is taken 0.3 s after that in (a).
(c-d) show a merging event of two clusters in a sample with $\phi=0.044\mbox{}\mu\mbox{m}^{-2}$.
Overlayed arrows show cell orientations. Two ellipses are drawn to
highlight difference in cell motility in/outside clusters. (e) Probability
density functions of averaged angular velocity of bacteria in(red)
/outside(green) clusters. (f) Schematic of the experimental setup
(not to scale). Scale bars in (a-b) and (c-d) correspond to 20 $\mu$m
and 15 $\mu$m, respectively. A coordinate frame is defined in (a)
and (f): the z axis points into the bulk fluid from the trapping (xy)
plane.}
\end{figure}

\textbf{\textit{Experiments}}\textit{ -} Our experiments are carried
out in drops of wild-type \textit{Serratia marcescens }(ATCC 274)
bacteria, which are propelled by a bundle of a few rotating flagella
\cite{Hesse2009}. For cultivation, small amount of bacteria from
frozen stock is put in 4 ml of Luria Broth (LB) growth medium consisting
of 0.5\% yeast extract (Sangon G0961), 1\% Tryptone (Sangon TN5250),
and 1\% NaCl (Sigma). Bacteria is incubated for 13 hours to a stationary
phase in a shaking incubator which operates at 30$^{\circ}$C and
200 rpm shaking speed. We then extract 1 ml bacteria solution and
re-grow bacteria in 10 ml fresh LB growth medium supplemented with
5$\mu$g/ml of A22 for 4 hours at 33$^{\circ}$C and 200 rpm. A22
is a small molecular inhibitor of a protein MreB, which is needed
to maintain rod-like shape of many bacteria \cite{White2010}. \textit{S.
marcescens} cells growing in a media with A22 are short ellipses with
a mean aspect ratio of 1.2. The final bacteria solution is further
diluted in distilled water to generate samples with various bacteria
densities. To control bacterial motility, 1$\mu$g/ml photosensitizer
FM 4-64 is added to the bacterial suspension \cite{Lu2013}. In the
presence of FM 4-64, the bacteria will be temporally paralyzed upon
exposure to strong light. A 100W Mercury lamp(Nikon C-SHG1) is used
to activate photodynamic effects. 

\begin{figure}
\centering{}\includegraphics[width=6.5cm]{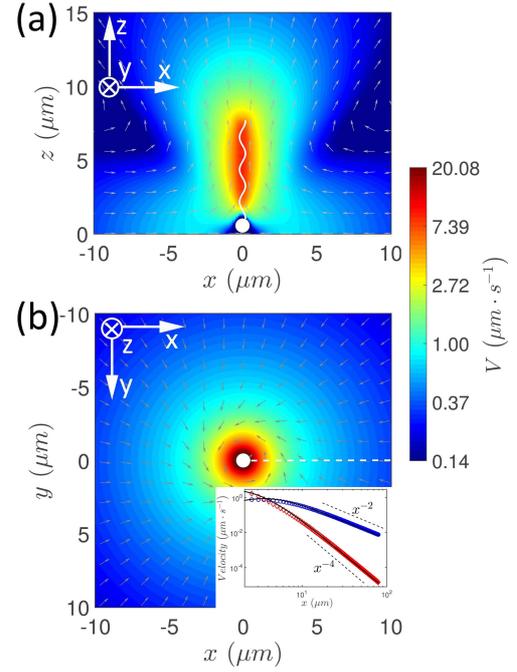}\protect\caption{(color online). Computed flow field around a model bacterium on two
planes: y = 0 in (a) and z = 0.54 $\mu$m in (b). Color represents
magnitude of velocity projection in the plane and arrows denote the
direction of flow. Profiles of radial (blue) and tangential (red)
velocity components along the white dashed line in (b) are plotted
as symbols in the insert with fits (lines) to analytical expressions
derived in Supplementary Text ST.pdf. Computation is carried out with
the Regularized Stockslet method \cite{Cortez2005,Rodenborn2013}
and details can be found in Supplementary Text ST.pdf.}
\end{figure}

Sample is enclosed in a sealed chamber (cf. Fig. 1(f)) which consists
of a silicone well, a plastic spacer, and two cover slips (0.13 mm
thick). The well has a diameter of 0.8 cm and a height of 0.1 cm.
In some experiments, 1 $\mu$m super-paramagnetic tracer beads are
added for flow visualization. Beads are confined to the interface
by a permanent magnet and the strength of confinement can be tuned
by moving the magnet relative to the sample. Two-dimensional bacterial
or tracer motion in the trapping plane is imaged through a 60X phase
contrast objective (Nikon ELWD ADL 60XC) and recorded by a camera
(Basler acA2040-180km). We use a holographic microscope to measure
three-dimensional motion of tracer particles and bacteria. A red LED
is used for illumination and recorded holograms are analyzed by the
Rayleigh-Sommerfeld back-propagation method to extract the spatial
coordinates of the scatter \cite{Sheng2006,Lee2007}.

\begin{figure*}
\centering{}\includegraphics[width=14cm]{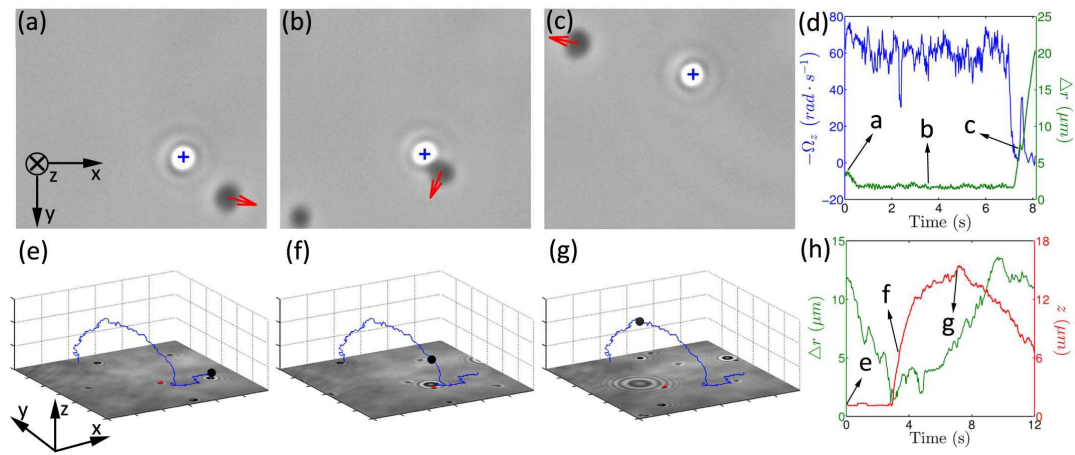}\protect\caption{(color online). (a-c) Interaction between a tracer bead (blue +) and
a bacterium in the trapping plane. The red arrows mark the instantaneous
orientation of cell body. (d) Time series of angular velocity (blue)
of the cell and the separation (green) between the cell and bead.
(e-f) Three-dimensional flow around a bacterium (red {*}) visualized
by a tracer bead (black sphere). In each panel (e-g), the black-and-white
image in xy plane is the raw hologram and the blue line is the tracer
trajectory. (h) Time series of the z-coordinate of the tracer (red)
and the separation (green) between the bead and the cell. Arrows in
(d) and (h) mark when data in other panels are recorded. \label{fig:3}}
\end{figure*}

\textbf{\textit{Bacteria clusters}}\textit{ -} \textit{S. marcescens}
is known to adhere strongly to the air-liquid interface possibly due
to hydrophobic interactions \cite{Syzdek1985,HEJAZIReview,Rabani2013}.
We use a holographic microscope to record such attaching events. Supplementary
movie S1.mp4 %
\footnote{The movie S1.mp4 shows the bacterium trajectory as a blue line and
the raw hologram as a black-and-white image. A black dot marks the
instantaneous position of the bacterium. The objective is focused
on the interface. All submitted movies are compressed in Xvid Codec and movies in MJPG format 
are available at:\newline \url{http://ins.sjtu.edu.cn/people/hpzhang/EPL/Movies.rar}%
} shows two typical events: bacteria swim from the bulk towards the
interface and get trapped near the interface. Trapped bacteria can
move freely in the trapping (xy) plane; their centers-of-mass are
tracked to quantify translational motion. Rotational motion is quantified
by following the principal axes of the elliptical bacteria. The main
control parameter in experiments is bacteria density in the trapping
plane which is quantified by the number of bacteria per unit area
and denoted as $\phi$. 

As shown in Fig. 1(a-b), trapped bacteria aggregate to form dynamic
clusters which are identified through Voronoi analysis of bacterial
positions (see Supplementary Text ST.pdf %
\footnote{ST.pdf contains detailed information of data analysis procedures and
of fluid dynamic analysis and is available at:\newline \url{http://ins.sjtu.edu.cn/people/hpzhang/EPL/ST.pdf}.%
} for details). Clustering formation is a robust phenomenon and occurs
under a wide range of bacteria densities and in various liquid environments
including LB media, motility buffer, and water. However, formation
of dynamic clusters requires bacterial motility. As shown in S2.mp4
\footnote{The movie S2.mp4 shows identified clusters in colors on phase-contrast
images. Light irradiation starts at 16.8 s in the video.%
}, clusters dissolve immediately after we reduce the motility by using
strong light \cite{Lu2013}. Clusters reappear when bacteria motility
is partially recovered after the light irradiation is stopped. Heavy
metal ions ($\mbox{CuSO}_{4}$) are also used to reduce motility\cite{Behkam2007};
similar results are obtained.

To understand the connection between motility and cluster formation,
we zoom-in and investigate how bacteria move in a cluster. We mark
a bacterium in a cluster with a red ellipse in Fig. 1(c-d) (cf. S3.mp4
\footnote{In the movie S3.mp4, phase-contrast images are shown in the background,
cell orientations are marked by arrows, and ellipses are used to highlight
two bacteria.%
}); the marked bacterium rotates its body in the trapping plane with
an angular velocity of $\Omega_{z}=-20$ rad/s. In contrast, a bacterium
outside cluster (marked in green) shows little change in its body
orientation. In Fig. 1(e) we plot the probability distribution functions
of angular velocity, $P\mbox{\ensuremath{\left(\Omega_{z}\right)}}$,
for bacteria in and outside clusters. While $P\mbox{\ensuremath{\left(\Omega_{z}\right)}}$
peaks around zero for bacteria outside clusters, $P\mbox{\ensuremath{\left(\Omega_{z}\right)}}$
shows a second peak at -23 rad/s for bacteria in clusters. We next
use fluid dynamic calculation and flow visualization to show that
the bacteria with large $\Omega_{z}$ can form clusters through hydrodynamic
interactions.

\begin{figure}
\centering{}\includegraphics[width=6cm]{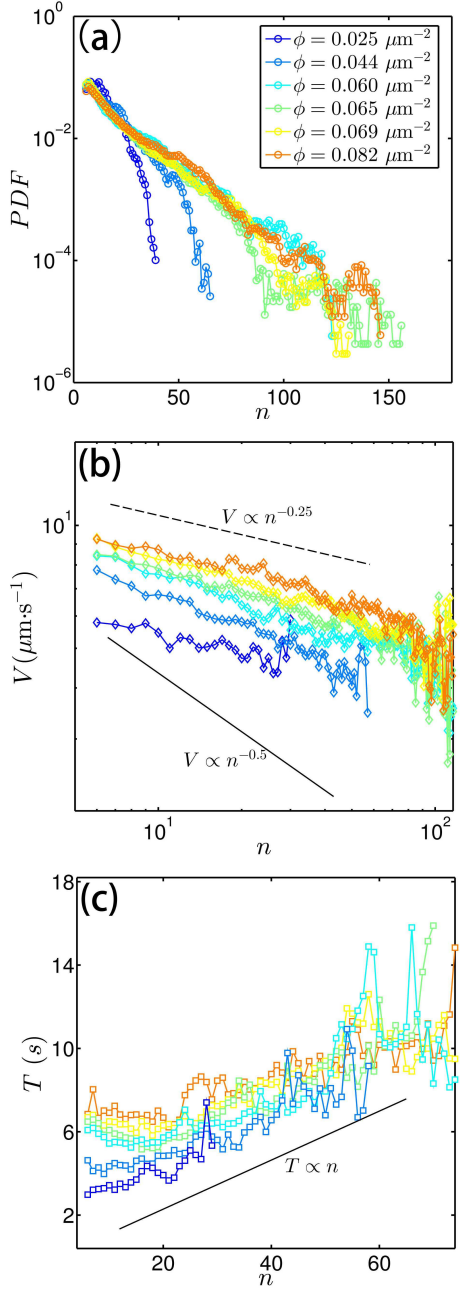}\protect\caption{(color online). Statistic and dynamic properties of clusters measured
for six different bacterial densities (color-coded according to the
legend in (a)). Quantities measured include probability distribution
function (a), mean translation speed (b) and mean rotation period
(c). Solid lines in (b) and (c) are salings derived from force and
torque balance (see text).\label{fig:4}}
\end{figure}

\textbf{\textit{Hydrodynamic interactions}}\textit{ -} \textit{S.
marcescens} bacteria swim by rotating their flagellar bundles. When
viewed from the front (body) of a bacterium, the bundle rotates clockwisely
and the cell body rotates in the opposite direction to achieve hydrodynamic
torque balance. Fast body rotation in Fig. 1(e) suggests the bacteria
in clusters orient their bundles perpendicular to the interface%
\footnote{A hydrodynamic mechanism was proposed to explain perpendicular orientation
of \textit{T. majus} cells \cite{Petroff}. Such a mechanism only
works at a no-slip boundary and doesn't apply in the case of \textit{S.
marcescens} near a liquid-air interface. %
}. Bacteria in such a configuration can generate fluid flow that leads
to cluster formation \cite{Roper2013,Pepper2013,Petroff}. To illustrate
the mechanism, we numerically compute fluid flow around a bacterial
model that is oriented perpendicular to the interface. As shown in
Fig. 2(a), the model has a 1-$\mu$m-diameter spherical body whose
center is located at (0, 0, 0.54 $\mu$m) and is driven by a rotating
flagellum. The translational degrees of freedom are frozen for the
model. The model bacterium exerts a force in the positive z direction
on the fluid, draws in fluid along the interface (at $z$ = 0) and
pushes fluid to the bulk. Rotation of the flagellum and cell body
also produces a tangential flow component that can be seen in Fig.
2(b), especially in the region close to the cell body. In the far
field, radial ($V_{r}\left(r\right)$) and tangential ($V_{\theta}\left(r\right)$)
components decays as: $V_{r}\left(r\right)\propto r^{-2}$ and $V_{\theta}\left(r\right)\propto r^{-4}$,
as shown in the insert. Bacteria outside clusters in our experiments
likely orient their bundle parallel to the interface and the cell
body shows little rotation perpendicular to the interface, i.e. $\Omega_{z}$
is small \cite{Zhang2010a,DiLeonardo2011,Chen2012}.

Important flow features in Fig. 2 are qualitatively confirmed in experiments
with two types of tracer beads. The first kind is strongly confined
in the z-direction and only probes fluid flow in the trapping plane.
A typical result is shown in Fig. 3(a-d) and Movie S4.mp4 %
\footnote{In the movie S4.mp4, a passive tracer confined to the interface visualizes
attractive flow around a rotating bacterium. %
}. A tracer bead is drawn to the bacterium from $t$ = 0 to $t$ =
0.5 s, which demonstrates the inward radial flow. While the bead and
bacterium are bound, they rotate around each other counterclockwisely,
which is a manifestation of the tangential flow in Fig. 2(b). At $t$
= 7 s, the bacterium stops rotating and swims away from the bead.
In the second experiment, a tracer bead is subject to weak confinement
and can be advected also in the z-direction. As shown in Fig. 2(e-h)
and Movie S5.mp4 %
\footnote{The Movie S5.mp4 shows three-dimensional motion of a passive tracer
around a rotating bacterium. %
}, the bead is drawn towards the bacterium in the trapping plane until
$t$ = 3.5 s, when the separation between the bead and bacterium is
about 1 $\mu$m. The bead is then advected quickly into the bulk,
which demonstrates strong flow into the bulk around the bacterium. 

Tracer particles are also used to visualize flow around bacteria clusters.
Movie S6.mp4 %
\footnote{Movie 6 shows three-dimensional motion of a passive tracer around
bacterial clusters. A 1$\mu$m tracer bead (marked by red$+$) is
drawn towards a bacterial cluster during the first 1.7s, and then
advected quickly into the bulk, as shown by the enlarging interference
rings in the hologram.%
} shows that clusters attract tracers and advect them into the bulk,
mimicking results in Figs. 2 and 3. This supports the following picture:
bacteria in clusters orient their flagella perpendicular to the interface;
they generate inward radial flow that attracts neighbors to form clusters
and counterclockwise tangential flow that drives clusters into rotation. 

\textbf{\textit{Cluster properties}}\textit{ -} Bacteria clusters
observed in experiments are highly dynamic; they constantly evolve
and change their sizes. We defined cluster size as the number of the
constituent bacteria, $n$. Probability distribution functions for
finding a cluster of a given size at six bacteria densities $\phi$
are shown in Fig. 4(a) . As the density increases, probability to
find large clusters increases. Probability distribution function decays
exponentially for large $n$. Similar exponential distributions have
been observed in many previous studies \cite{Zhang2010a,Chen2012}
and may be modeled by fusion-fission processes \cite{Gueron1995}. 

We compute the following quantities to quantify translation and rotation
of the $I$th cluster which contains the $i$th bacterium at a location
$\vec{r}_{i,I}$ and with a velocity $\vec{v}_{i,I}$. The center-of-mass
of the $I$th cluster is located at $\vec{R}_{I}=\left\langle \vec{r}_{i,I}\right\rangle _{i}$,
where $\left\langle \cdot\right\rangle _{i}$ denotes an average over
all $n_{I}$ bacteria in the $I$th cluster. Speed of the center-of-mass
is $V_{I}=\left|\vec{V}_{I}\right|=\left|\left\langle \vec{v}_{i,I}\right\rangle _{i}\right|.$
The angular speed of the $I\mbox{th}$ cluster is defined as: $\omega_{I}=\left|\frac{\left\langle \left(\vec{r}_{i,I}-\vec{R}_{I}\right)\times\left(\vec{v}_{i,I}-\vec{V}_{I}\right)\right\rangle _{i}}{\left\langle \left(\vec{r}_{i,I}-\vec{R}_{I}\right)\cdot\left(\vec{r}_{i,I}-\vec{R}_{I}\right)\right\rangle _{i}}\right|,$
whose inversion is the rotation period: $T_{I}=\frac{1}{\omega_{I}}.$
Averaging $V_{I}$ and $T_{I}$ over all clusters of size $n$, we
have mean translation speed $V\left(n\right)=\left\langle V_{I}\right\rangle _{n_{I}=n}$
and mean rotation period $T\left(n\right)=\left\langle T_{I}\right\rangle _{n_{I}=n}$
. 

Results in Fig. 4 (b-c) show that, for a given $\phi$, larger clusters
translate and rotate slower than smaller ones. This dependence is
qualitatively consistent with a simple model of active clusters made
of self-propelled particles \cite{Petroff}. Each particle is driven
by a propulsive force $F_{sp}$ in the $\hat{e}_{i}$ direction and
has a drag coefficient of $\gamma$; they interact through pair-wise
forces in the radial and tangential directions . If $n$ particles
in a cluster are randomly oriented, total propulsive force on the
cluster is $\left|\sum_{i}F_{sp}\hat{e}_{i}\right|=F_{sp}n^{0.5}$.
The total friction coefficient is the sum of all particles: $n\gamma$.
The average velocity of the center-of-mass is the ratio of the total
force to the total friction coefficient and scales as: $V\propto n^{-0.5}$.
Using a similar torque-balance argument \cite{Petroff,Yan2015}, we
can get a scaling law for rotation period: $T\propto n$ . These two
scalings are shown in Fig. (b) and (c) as black lines; discrepancies
with experimental results can be clearly seen.

First, we notice that cluster motion in experiments depends strongly
on global densities, $\phi$. Fig. 4 (b-c) show that a cluster of
a given size translates faster and rotates slower in a system with
a higher density. The dependence on system density likely arises from
long-ranged hydrodynamic interactions which enable fluid disturbances
to propagate far and couple different clusters over large distances.
Second, Fig. 4(b) shows that cluster velocity $V\left(n\right)$ scales
as $V\propto n^{-0.25}$ rather than $V\propto n^{-0.5}$. This is
also possibly related to hydrodynamic interactions which determine
how cluster friction and net propulsive force scale with cluster size.
In the limit of Stokes drag, cluster friction scales as $\sqrt{n}$
\cite{Cremer2014} in two dimensions, which suggests that net propulsive
force of clusters may scale as $n^{0.25}$. Data in Fig. 4(b-c) indicate
that hydrodynamic interactions in our experiments are too complex
to be represented by simplified pair-wise forces,.

\textbf{\textit{Discussion -}} Inward radial flow in Fig. 2(a) is
similar to feeding flow that many micro-organisms use to gather food
from the fluid environment near an interface \cite{Roper2013,Pepper2013}.
The same flow pattern is also used to explain the formation of bound
\textit{Volvox} pair \cite{Drescher2009} and to explain the attractive
force between thermophoretic colloids \cite{Weinert2008,DiLeonardo2009,Yang2013}.
Such an attractive boundary flow should exist near any low-Reynolds
number swimmer that is oriented perpendicular to a fluid or solid
boundary and swims into the boundary; consequently, these oriented
swimmers experience effective attraction and can be hydrodynamically
assembled into clusters. This provides a new mechanism, besides phoretic
\cite{Palacci2013}, depletion \cite{Schwarz-Linek2012} interactions
and self-trapping effects \cite{Buttinoni2013}, to generate clusters
of active particles. 

Petroff \textit{et al} \cite{Petroff} recently reported that \textit{T.
majus }bacteria form two-dimensional crystals near a liquid-solid
interface through a mechanism similar to that in \textit{S. marcescens
}clusters. However, \textit{T. majus} and \textit{S. marcescens}
systems are significantly different in at least two aspects. First,
the strength of attractive interaction is different. Petroff \textit{et
al} used the product of propulsive force and swimmer size to estimate
the energy scale (denoted as $E$) for attraction between cells.\textit{
T. majus }have an average diameter of 8.5 $\mu\mbox{m}$ and swim
at a speed of 600 $\mu\mbox{m}/s$; it was found that the attractive
energy is much larger than thermal energy: $E\sim10^{4}k_{B}T$. Consequently,
\textit{T. majus} crystals are very stable and can contain up to a
thousand cells. On the other hand, \textit{S. marcescens}, like many
other commonly studied bacteria \cite{Sokolov2007,Zhang2009,Schwarz-Linek2012}
and artificial swimmers\cite{Palacci2013,Buttinoni2013}, are approximately
ten times smaller in both size and speed than \textit{T. majus}; the
attractive energy scale is 1000 times smaller and is on the order
of $10k_{B}T$. Therefore, fluctuations play a more important role
in the \textit{S. marcescens }system and render \textit{S. marcescens}
clusters dynamic and constantly evolving. Second, \textit{T. majus}
crystals and \textit{S. marcescens} clusters form near different hydrodynamic
boundaries. A no-slip boundary makes the hydrodynamic attraction,
$f$, between two \textit{T. majus} cells decays rapidly as their
separation, $r$, increases: $f\sim r^{-4}$; for \textit{S. marcescens}
clusters near a free-slip boundary, we have $f\sim r^{-2}$. Consequently,
while hydrodynamic interaction between \textit{T. majus} cells are
severely screened, Fig. 4 shows that long-ranged hydrodynamic interactions
influence \textit{S. marcescens} clusters properties. 

\textbf{\textit{Conclusion - }}In summary, we have investigated dynamic
clusters of \textit{S. marcescens} bacteria near an air-liquid interface.
Fluid dynamic calculation and flow visualization suggest that the
constituent bacteria of these clusters orient their flagella perpendicular
to the interface. Bacteria in such a configuration generate radial
flow that attracts neighbors to form clusters and tangential flow
that sets clusters into counter-clockwise rotation. We measured statistical
properties of bacteria clusters and showed cluster properties are
affected by long-ranged hydrodynamic interactions. \textit{S. marcescens}
clusters efficiently change material and fluid transport near the
air-liquid interface; they may have environmental and biological consequences
\cite{Dombrowski2004}. 

\acknowledgments We acknowledge financial supports of the NSFC (No.
11422427, No. 11404379), the Program for Professor of Special Appointment
at Shanghai Institutions of Higher Learning (No. SHDP201301), and
the Innovation Program of Shanghai Municipal Education Commission
(No. 14ZZ030).

\bigskip{}

\end{document}